\documentclass[aps,prl,twocolumn,showpacs,superscriptaddress]{revtex4-1}

\usepackage[dvipdfm]{graphicx}
\usepackage[dvipdfm]{hyperref}
\usepackage{bm}

\begin{document}

\title{Direct measurement of the optical trap-induced decoherence}

\author{Nobuyuki Matsumoto}
\email{nobuyuki.matsumoto.e7@tohoku.ac.jp}
  \affiliation{Frontier Research Institute for Interdisciplinary Sciences, Tohoku University, Sendai 980-8578, Japan}
  \affiliation{Research Institute of Electrical Communication, Tohoku University, Sendai 980-8577, Japan}
  \affiliation{JST, PRESTO, Kawaguchi, Saitama 332-0012, Japan}
\author{Kentaro Komori}
 \affiliation{Department of Physics, University of Tokyo, Bunkyo, Tokyo 113-0033, Japan}
\author{Sosuke Ito}
\affiliation{Department of Physics, Tokyo Institute of Technology, Meguro, Tokyo 152-8551, Japan}
\author{Yuta Michimura}
\affiliation{Department of Physics, University of Tokyo, Bunkyo, Tokyo 113-0033, Japan}
\author{Yoichi Aso}
  \affiliation{National Astronomical Observatory of Japan, Mitaka, Tokyo 181-8588, Japan}
 \affiliation{Department of Astronomical Science,  SOKENDAI (The Graduate University
for Advanced Studies), Mitaka, Tokyo 181-8588, Japan}

\date{\today}

\begin{abstract}
Thermal decoherence is a major obstacle to the realization of quantum coherence for massive mechanical oscillators. 
Although optical trapping has been used to reduce the thermal decoherence rate for such oscillators, it also  increases the rate by subjecting the oscillator to stochastic forces resulting from the frequency fluctuations of the optical field, thereby setting a fundamental limit on the reduction. 
This is analogous to the noise penalty in an active feedback system. 
Here, we directly measure the rethermalizaton process for an initially cooled and optically trapped suspended mirror, and identify the current limiting decoherence rate as due to the optical trap. 
Our experimental study of the trap-induced decoherence rate will enable future advances in the probing of fundamental quantum mechanics in the bad cavity regime, such as testing of deformed commutators. 
\end{abstract}


\maketitle

{\it Introduction.}---
Various types of optical potentials have been used to change the dynamics of mechanical systems, including atoms, thin membranes and suspended mirrors in order to, e.g. observe signatures of shot-noise radiation-pressure fluctuations \cite{murch2008}, enhance the quality factor of the system \cite{norte2012}, and improve the sensitivity of gravitational-wave detectors \cite{braginsky1997,miyakawa2006,buonanno2002,somiya2016}. 
Since an optical potential works as an ideal spring for the trapped mode in terms of energy dissipation, it can reduce the number of quanta in the mode (the so-called ``optical dilution" effect) so that even a low-frequency massive oscillator will exhibit quantum behavior \cite{braginsky1999}. 
Progress towards this quantum regime is underway \cite{sheard2004,corbitt2007-1,corbitt2007,norte2012} in the field of cavity optomechanics \cite{aspelmeyer2014} particularly in the bad cavity regime, where the optical linewidth is broader than that of the mechanical resonance. 
Although the bad cavity condition is often not promising in terms of coherence because of the slow mechanical oscillation, it enables us to detect gravitational-waves \cite{novel2016}, and potentially to probe deformed commutators \cite{pikovski2012,bawaj2015}, generate entangled states \cite{marshall2003,pepper2012,muller2008}, and test wavefunction collapse models \cite{miao2010,Bahrami2014,nimmrichter2014,diosi2015,li2016}. 

The reported limits of optical dilution so far are due to position sensing noise resulting from frequency fluctuations of the laser \cite{corbitt2007}, 
 the structural effect of the mechanical oscillator \cite{norte2012}, and also parametric instability \cite{kippenberg2005,miyakawa2006,corbitt2006,arcizet2006-1}. 
Additionally, the reduction of thermal decoherence rate (i.e. the derivative of occupation of the mode with respect to the time) has been observed indirectly by measuring the stationary state of the trapped mode \cite{corbitt2007}. 
However, at the same time, the optical potential induces heating of the oscillator so that it increases decoherence, which is analogous to the ``noise penalty" in an active feedback system \cite{taubman1995}. 
The excess decoherence arises because the passive feedback loop (i.e. the optical potential) subjects the oscillator to stochastic forces resulting from frequency fluctuations of the laser. 
This phenomenon sets a fundamental limit on the reduction of decoherence even in the absence of active feedback. 

In this paper, we report an experimental study of optical trap-induced decoherence. 
We cool an optically trapped pendulum mode (suspended mirror) based on measurement of the pendulum's displacement  (cf. \cite{poggio2007,mancini1998,cohadon1999,arcizet2006,wilson2015}). 
By turning off the cooling, we measure the time evolution of the initially cooled mode, and hence the thermal decoherence rate. 
We measure this rate as the optical rigidity is varied. 
We find that the reduction of decoherence is limited by the optical-trap induced decoherence, whose effect is proportional to the frequency noise spectrum $S_{\dot{\phi}}$ at the resonance $\omega_{\rm eff}$ of the trapped mode. 
For an optically trapped oscillator, the condition
\begin{eqnarray}
S_{\dot{\phi}}(\omega_{\rm eff})<\frac{g_0^2}{\omega_{\rm eff}}
\label{condition}
\end{eqnarray}
must be satisfied in order to measure the quantum coherence of the oscillator, where $g_0$ is the optomechanical coupling per single photon. 
Since $\omega_{\rm eff}$ must be larger than the decoherence rate, this is more stringent than the requirement for achieving ground-state cooling \cite{rabl2009}. 
Condition (\ref{condition}) is experimentally challenging in the case of using a massive mechanical oscillator, but we show that it is feasible in the presence of frequency stabilization. 

\begin{figure*}
\centering
\includegraphics[width=59mm]{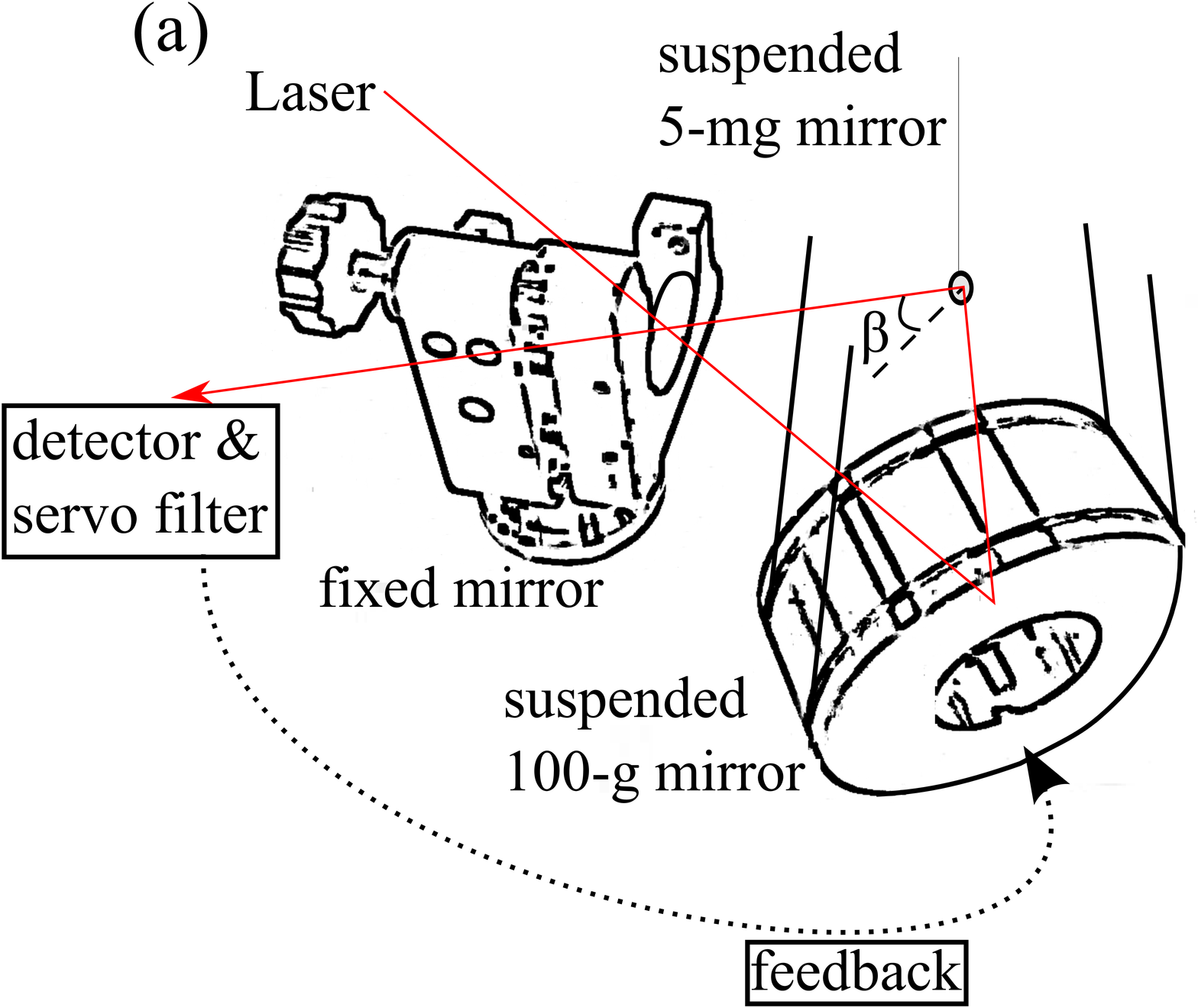}
\includegraphics[width=59mm]{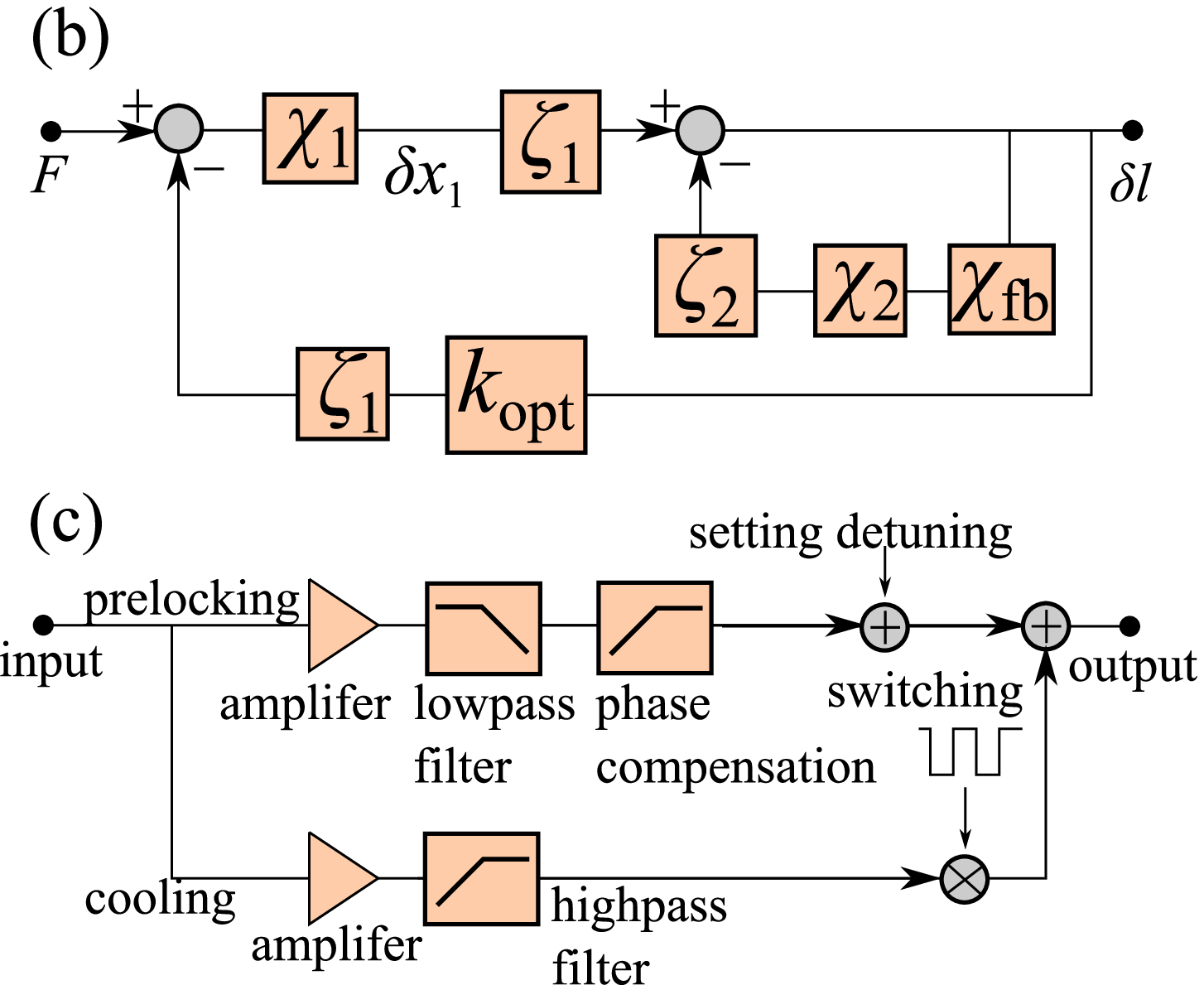}
\includegraphics[width=59mm]{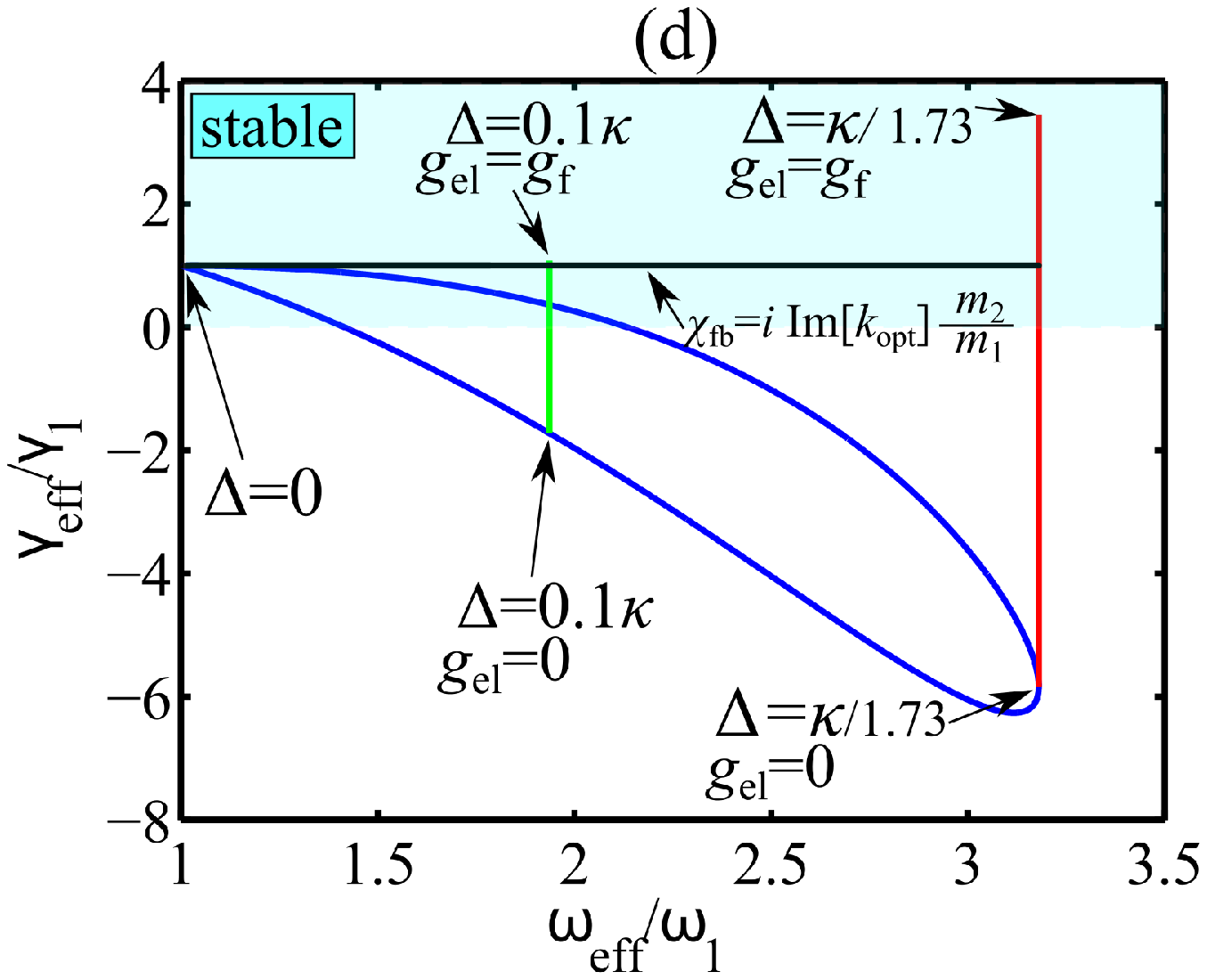}
\caption{(Color online) (a) Experimental setup. 
We focus on the pendulum mode of the 5-mg movable mirror. 
(b) The corresponding block diagram. 
Minus-signs around the gray disks represent negative feedback. 
$\delta l$ is the displacement of the cavity. 
(c) The servo filters. 
After prelocking of the cavity length, it is switched off and then the gain of the highpass filter is increased. 
By mixing square waves, the cooling system is rapidly switched off so that the thermal decoherence rate can be observed. 
(d) Parametric plot of the optical spring (as a curve) and optoelectrical feedback (as vertical lines and a horizontal line) as a function of detuning $\Delta$ and gain $g_{\rm el}$ of the feedback system $\chi_{\rm fb}=i\omega g_{\rm el}$, respectively. 
At zero detuning from resonance ($\Delta=0$), both the resonant frequency and the damping rate are equal to the intrinsic ones. 
At zero gain of the electrical servo ($g_{\rm el}=0$), feedback cooling vanishes. 
Parameter values are $\omega_{\rm 1}/(2\pi)=\omega_{\rm 2}/(2\pi)=1\ {\rm Hz}$, $\gamma_{\rm 1}/(2\pi)=10^{-6}\ {\rm Hz}$, $\gamma_{\rm 2}/(2\pi)=10^{-2}\ {\rm Hz}$, $\kappa/(2\pi)=\kappa_{\rm in}/(2\pi)=2\times10^6\  {\rm Hz}$, $g=\omega_{\rm laser}$, $\bar{n}_{\rm cav}=8.5\times10^5/[1+(\Delta/\kappa)^2]$, and $g_{\rm f}=2\times10^{-8}\ {\rm N/m/Hz}$.}
\label{fig1}
\end{figure*}

{\it Experimental design.}---
We use an optical cavity consisting of a suspended 5-mg(=$m_{\rm 1}$) mirror, a suspended $1\times10^2$-g(=$m_{\rm 2}$) mirror which is attached to coil-magnet actuators for active feedback, and a fixed mirror [Fig. \ref{fig1}(a)]. 
This design makes the system stable with respect to the mirror's yaw motion \cite{matsumoto2014}. 
The cavity is pumped by an optical laser blue-detuned from the cavity frequency $\omega_{\rm cav}/(2\pi)$ so that the mirrors are coupled to an optical spring with spring constant $k_{\rm opt}$. 
The cavity length fluctuation is obtained by monitoring the reflected light by a photo detector, whose voltage signal is filtered and then fed back to the actuators. 
The dynamics of the cavity optomechanical system shown in Fig. \ref{fig1}(a) can be modeled as a linear (negative) feedback system and the corresponding block diagram is shown in Fig. \ref{fig1}(b). 
The mechanical susceptibility (i.e. displacement response to an applied force $F$) of the 5-mg movable mirror can be determined by Mason's rule \cite{mason1956}:
\begin{eqnarray}
\chi_{\rm eff}\equiv\frac{\delta x_{\rm 1}}{F}=\chi_{\rm 1}\frac{1+\zeta_2\chi_{\rm 2}\chi_{\rm fb}}{1+\zeta_1^2\chi_{\rm 1}k_{\rm opt}+\zeta_2\chi_{\rm 2}\chi_{\rm fb}}
\label{eq2}
\end{eqnarray}
Here $\delta x_{\rm 1}$ is the displacement of the movable mirror, $\chi_{\rm 1} (\chi_{\rm 2})$ is the bare mechanical susceptibility of the 5-mg ($1\times10^2$-g) mirror, and $\zeta_1$ ($\zeta_2$) is the derivative of the cavity length $l$ with respect to the position of the the 5-mg ($1\times10^2$-g) mirror. 
$\chi_{\rm fb}$ consists of the derivative of the cavity's power reflectivity with respect to the cavity length, the efficiency of the photo detector, the servo circuits [details are shown in Fig. \ref{fig1}(c)], and the actuation efficiency. 
The optical spring constant is given by the following equation in the frequency domain \cite{aspelmeyer2014}: 
\begin{eqnarray}
k_{\rm opt}=2\hbar g^2\bar{n}_{\rm cav} \frac{\Delta}{(\kappa+i\omega)^2+\Delta^2}
\label{eq5}
\end{eqnarray}
Here $g$ is the light-enhanced coupling rate for the linearized regime, $\bar{n}_{\rm cav}$ is the average photon number in the cavity, $\Delta$ is the cavity detuning, $\kappa$ is the cavity amplitude decay rate, $\hbar$ is the reduced Planck constant and $\omega/(2\pi)$ is the Fourier frequency. 
For the massive and the fixed mirrors, the optical spring is negligible since it hardly changes the dynamics of the these mirrors. 
From Eqs. (\ref{eq2}) and (\ref{eq5}), dissipation of the movable mirror is controlled by the (servo) highpass filter because damping forces made by the filter are transfered to the movable mirror through the optical rigidity. 
For the fixed feedback gain, the effective dissipation (i.e. degree of cooling) thus depends on the optical rigidity [cf. vertical lines in Fig. \ref{fig1}(d)]. 
For strong optical rigidity, the optical anti-damping is canceled by the following feedback system [the horizontal line in Fig. \ref{fig1}(d)]:
\begin{eqnarray}
\chi_{\rm fb}=i{\rm Im}[k_{\rm opt}] \frac{m_{\rm 2}}{m_{\rm 1}}\simeq-i\omega \frac{m_{\rm 2}\omega_{\rm eff}^2}{\kappa} 
\label{eq}
\end{eqnarray}
Here we suppose $\omega\ll\sqrt{\Delta^2+\kappa^2}$.
Roughly speaking, the system becomes unstable unless the feedback gain $g_{\rm el}$ exceeds $m_{\rm 2}\omega_{\rm eff}^2/\kappa$ because the bare dissipation rate is generally very small. 

Note that the optoelectrical cooling presented here is ``active'' because the servo system directly suppresses thermal motion. 
Passive cooling with a servo-modified optical spring has been reported by C. M. Mow-Lowry et al.  \cite{mow-lowry2008}. 
An advantage of active cooling is that it is possible to realize the ground state for the bad cavity condition given by $\omega_{\rm eff}\ll\kappa$ \cite{genes2008}. 
This condition is also required to probe spacetime using pulsed light \cite{pikovski2012}.\par

\begin{figure*}
\begin{minipage}{0.32\hsize}
\includegraphics[width=59mm]{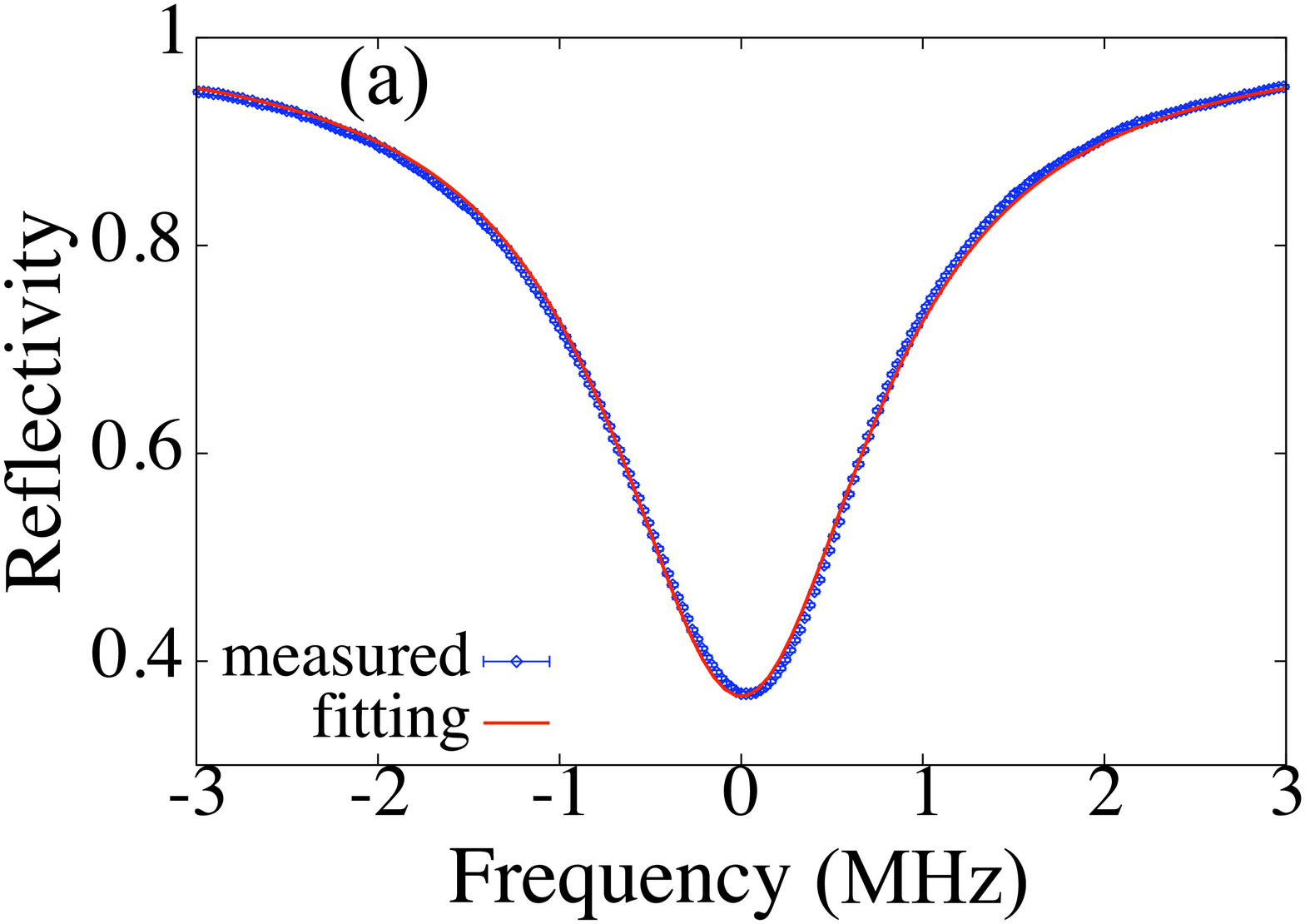}
\includegraphics[width=59mm]{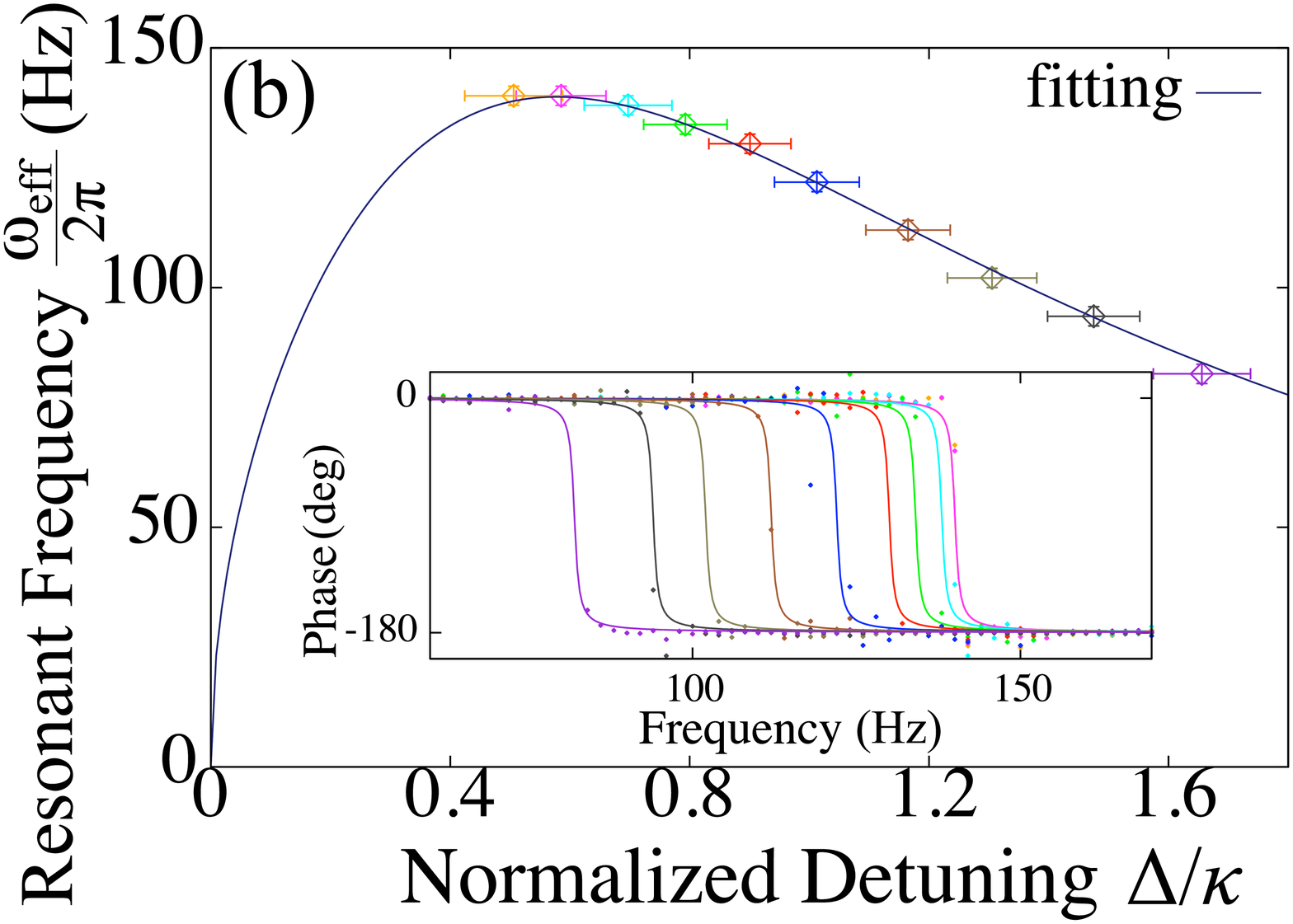}
\end{minipage}
\begin{minipage}{0.32\hsize}
\includegraphics[width=59mm]{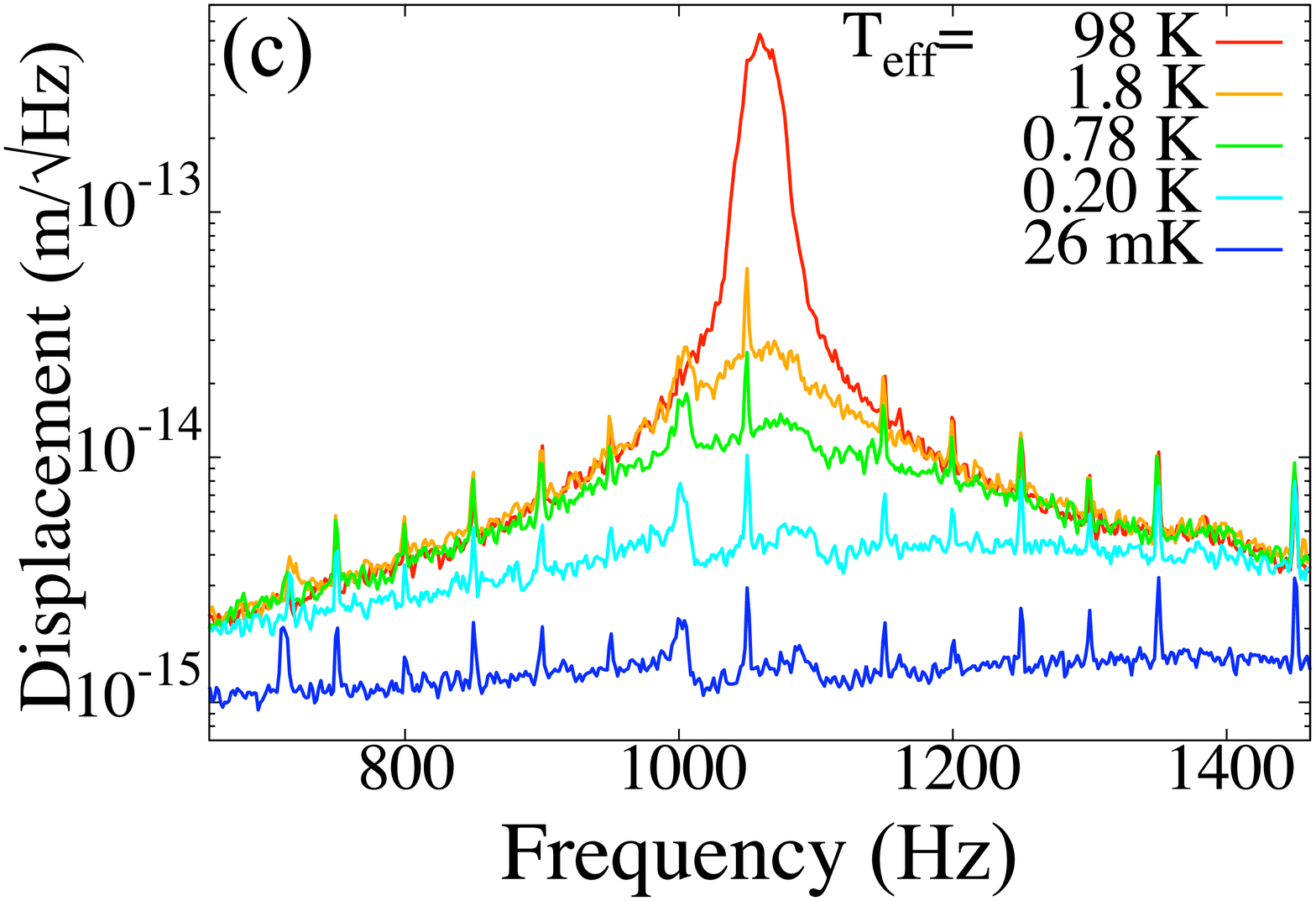}
\includegraphics[width=59mm]{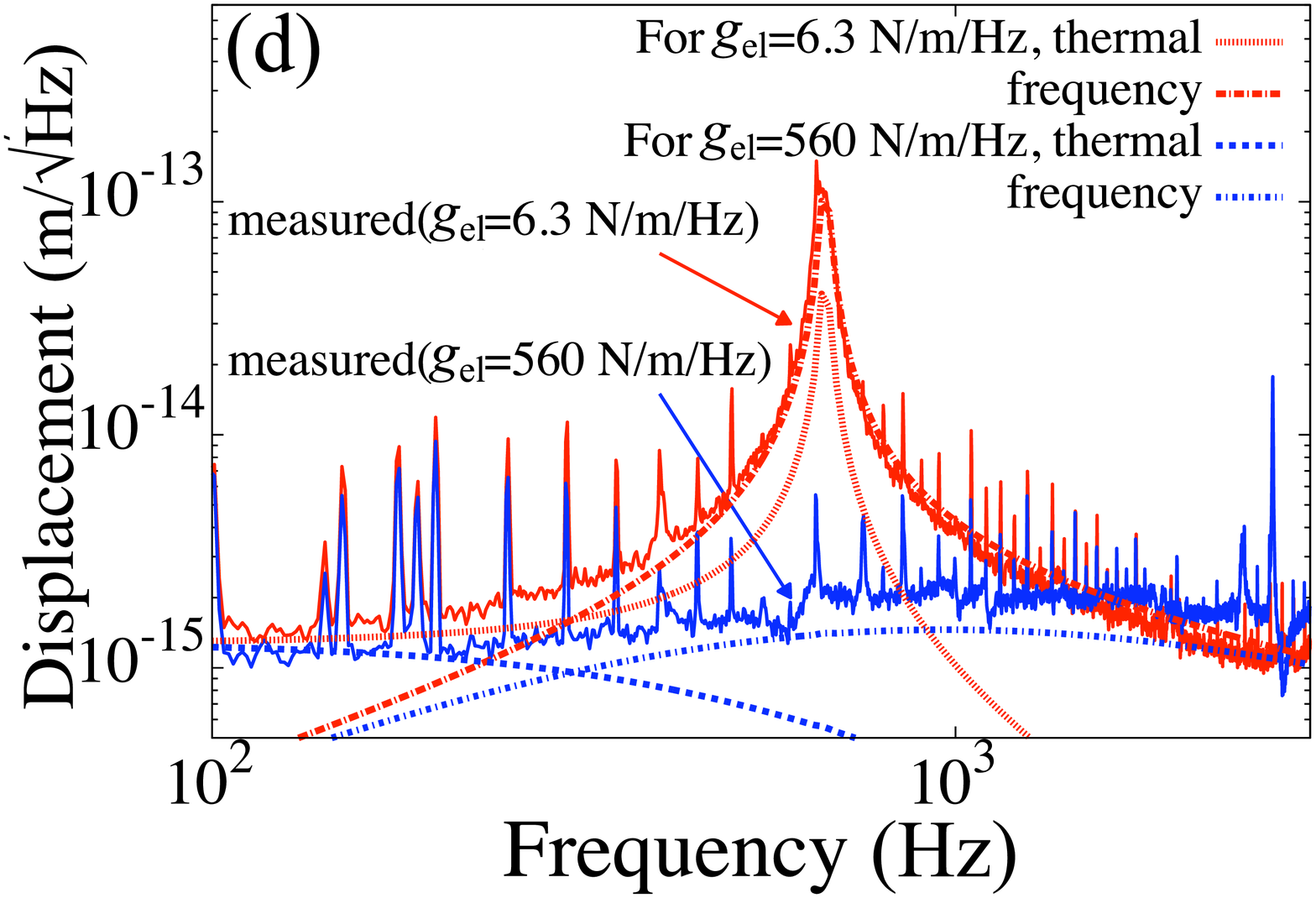}
\end{minipage}
\begin{minipage}{0.32\hsize}
\includegraphics[width=59mm]{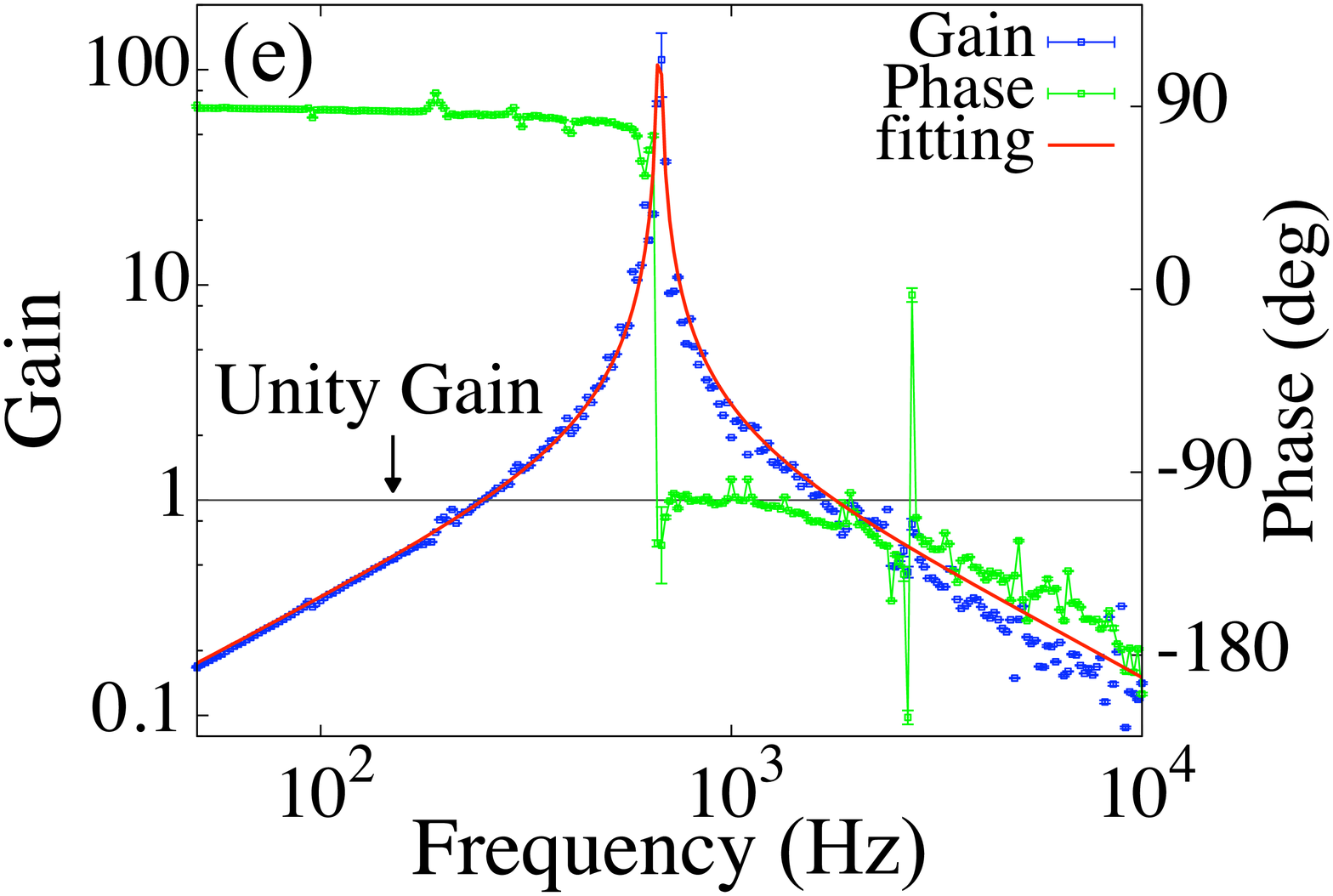}
\includegraphics[width=59mm]{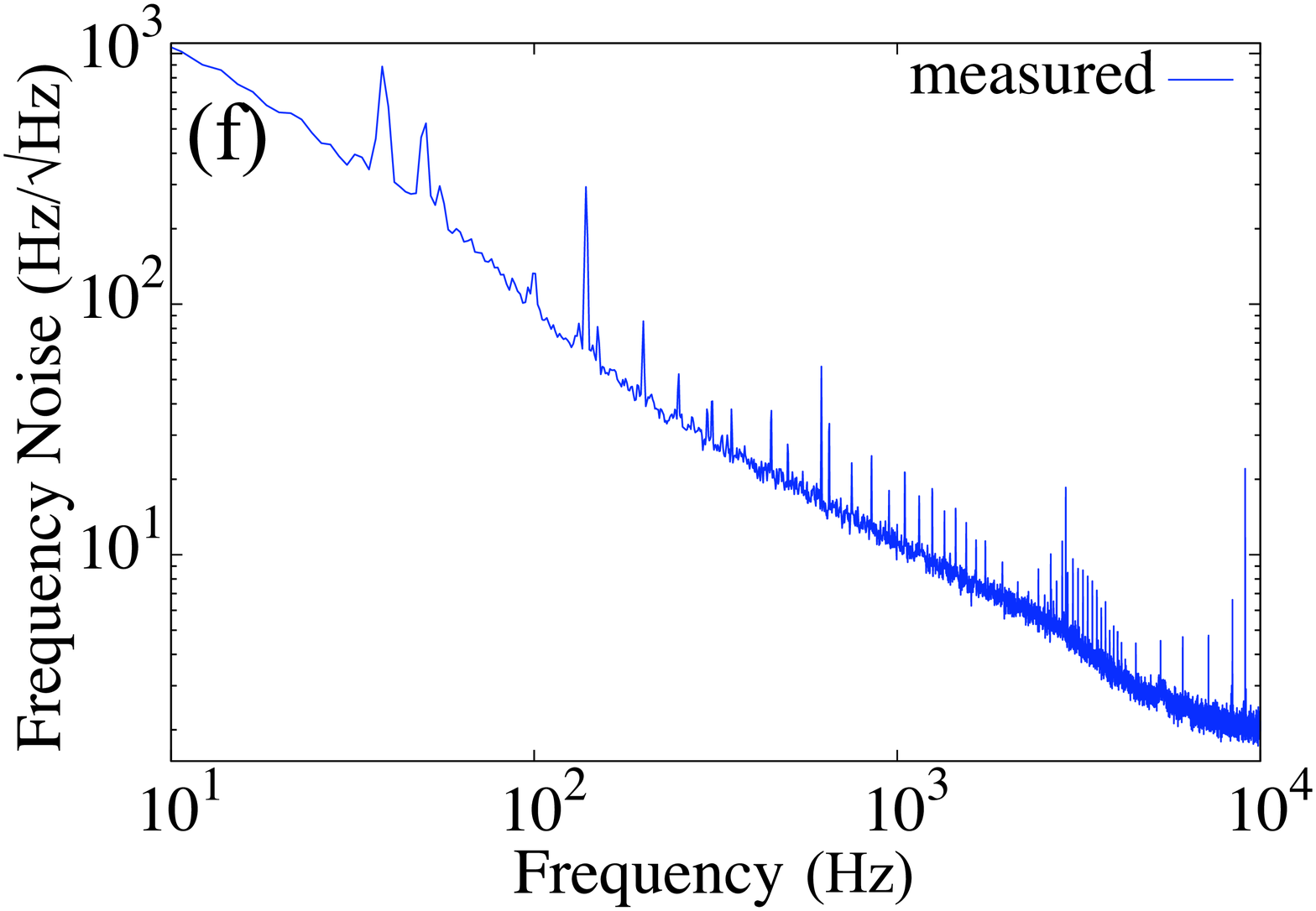}
\end{minipage}
\caption{(Color online) (a) Measured optical linewidth. 
(b) The response of the optical rigidity as a function of the cavity detuning. 
Inset shows the phase response. 
(c) and (d) For the case of $\omega_{\rm eff}/(2\pi)=1.06\ {\rm kHz},\ {\rm and}\ 662\ {\rm Hz}$. 
The measured spectra are shown as the solid curves. 
(e) For the case of $\omega_{\rm eff}/(2\pi)=662\ {\rm Hz}$, and $g_{\rm el}=560\ {\rm N/m/Hz}$. 
The measured open-loop transfer function of the cooling system is shown as squares. 
(f) Measured square root of the frequency noise spectrum by using a rigid reference cavity and the Pound-Drever-Hall (PDH) method \cite{drever1983}. 
}
\label{fig3}
\end{figure*}

{\it Parameters.}---
The cavity has a round-trip length $L=8.7\pm0.2\ {\rm cm}$ and is installed in a chamber at a pressure of 9 Pa. 
The 5-mg ($1\times10^2$-g) mirror has a bare pendulum frequency $\omega_{\rm 1}/(2\pi)$ ($\omega_{\rm 2}/(2\pi)$) of 2.14 Hz (2.89 Hz). 
The 5-mg mirror is suspended by a tungsten wire 3 um in diameter with a bare energy dissipation rate of $\gamma_{\rm 1}/(2\pi)=(1.2\pm0.1)\times10^{-2}\ {\rm Hz}$ [i.e. $Q_{\rm 1}\equiv\omega_{\rm 1}/\gamma_{\rm 1}=(1.8\pm0.1)\times10^2$], while the $1\times10^2$-g mirror is suspended by four tungsten wires 30 um in diameter with a bare rate of $\gamma_{\rm 2}/(2\pi)=(5.4\pm0.2)\times10^{-2}\ {\rm Hz}$ (i.e. $Q_{\rm 2}=54\pm2$). 
The finesse of the cavity is $\mathcal{F}=(1.98\pm0.08)\times10^3$ corresponding to a total amplitude decay rate of $\kappa/(2\pi)=0.84\pm0.03\ {\rm MHz}$. 
The ratio of the input coupler's decay rate to the total decay rate is $\kappa_{\rm in}/\kappa=0.19\pm0.01$. 
The laser frequency is $\omega_{\rm laser}/(2\pi)\simeq300\ {\rm THz}$ (Coherent Inc. Mephisto), and the maximum input laser power to the cavity is $47\pm5\ {\rm mW}$. 
The beam is incident on the movable mirror at an angle $\beta$, where $\cos\beta=0.78\pm0.04$. 

{\it Calibration of the pendulum's displacement.}--- To obtain the (one-sided) displacement spectrum $S_{\rm x}(\omega)$, we observe the reflected light from the cavity. 
For blue-detuning satisfying $\omega_{\rm eff}\gg\omega_{\rm 1}$, we obtain the following equation by Mason's rule \cite{mason1956} and Eqs. (\ref{eq2}) and (\ref{eq5}): 
\begin{eqnarray}
\sqrt{S_{\rm V}(\omega)}=\sqrt{S_{\rm x}(\omega)}\frac{2\pi c m_{\rm 1}}{\mathcal{F}\zeta_1}\left(1-\frac{\kappa_{\rm in}}{\kappa}\right)\omega_{\rm eff}^2\eta
\label{eq10}
\end{eqnarray}
Here $S_{\rm V}(\omega)$ is the spectrum of the observed voltage signal, $c$ is the speed of light, and $\eta$ is the voltage-to-power conversion factor of the detector. 
Note that this technique does not require us to measure the cavity detuning, which is hard accurately determine. 
To obtain $S_{\rm x}(\omega)$ from Eq. (\ref{eq10}), we first determine the finesse by sweeping the laser frequency over the cavity resonance [Fig. \ref{fig3}(a)]. 
We then trap the movable mirror using the optical spring created by the $0.82\pm0.08$ mW incident laser. 
We obtain the ratio $\kappa_{\rm in}/\kappa$ and $\zeta_1=2\cos{\beta}$ by measuring the dependence of the effective resonant frequency on the detuning [Fig. \ref{fig3}(b)]. \par

{\it Cooling.}--- After prelocking the cavity length, we make the gain equal to zero so that the length is passively controlled by the optical potential. 
We then increase the gain of the highpass filter to cool the trapped mode.\par

We show examples of the cooling spectra in Figs. \ref{fig3}(c)-(d) with varying the gain of the highpass filter. 
In Fig. \ref{fig3}(c) higher harmonics of the AC power supply frequency of 50 Hz are visible through the spectrum. 
We integrate each observed spectral peak within 3 $\sigma$ to obtain the mode's temperature $T_{\rm eff}=m_{\rm 1}\omega_{\rm eff}^2\langle x^2\rangle/k_{\rm B}$, where $k_{\rm B}$ is the Boltzmann's constant. 
The lowest temperature is $15\pm3$ mK at $\omega_{\rm eff}/(2\pi)=(6.62\pm0.07)\times10^2$ Hz. 
For the case of $\omega_{\rm eff}/(2\pi)>662$ Hz, the degree of cooling is saturated since resonance at 2.7 kHz [cf. Fig. \ref{fig3}(e)] prevents us from increasing the gain to a sufficiently high level. 

For the noise analysis, we find that the frequency fluctuations of the pump laser [characterized by the noise spectrum $S_{\rm \dot{\phi}}\ ({\rm Hz^2/Hz})$] and the pendulum's Brownian motion [thermal noise $S_x^{\rm th}\ ({\rm m^2/Hz})$]  dominate. 
Frequency fluctuations drive the oscillator through the optical rigidity, whose displacement spectrum is given, according to Mason's rule \cite{mason1956}, by $\sqrt{S_{x}^{\rm freq}(\omega)}=\sqrt{S_{\rm \dot{\phi}}}| \chi_{\rm eff}/[\chi_{\rm 1}(1+\zeta_2\chi_{\rm 2}\chi_{\rm fb})g]|$. 
We obtain $\chi_{\rm fb}$ by measuring the open-loop gain $\zeta_2\chi_{\rm 2}\chi_{\rm fb}/(1+\zeta_1^2\chi_{\rm 1}k_{\rm opt})$ [Fig. \ref{fig3}(e)]. 
Assuming $\sqrt{S_{\rm \dot{\phi}}(\omega)}=10\ {\rm kHz}/[\omega/(2\pi)]\ {\rm Hz/\sqrt{Hz}}$ based on the measured frequency noise spectrum shown in Fig. \ref{fig3}(f), we then obtain $\sqrt{S_x^{\rm freq}(\omega)}$ [dash-dotted lines in Fig. \ref{fig3}(d)]. 
The thermal noise spectrum is given, according to the fluctuation-dissipation theorem \cite{callen1952}, by $\sqrt{S_x^{\rm th}(\omega)}=\sqrt{4k_{\rm B}T\gamma_{\rm 1}m_{\rm 1}}|\chi_{\rm eff}|$. 
For the case of $\omega_{\rm eff}/(2\pi)= 662$ Hz, the thermal noise dominates under the resonance [dotted lines in Fig. \ref{fig3}(d)]. 

\begin{figure}
  \centering
\includegraphics[width=80mm]{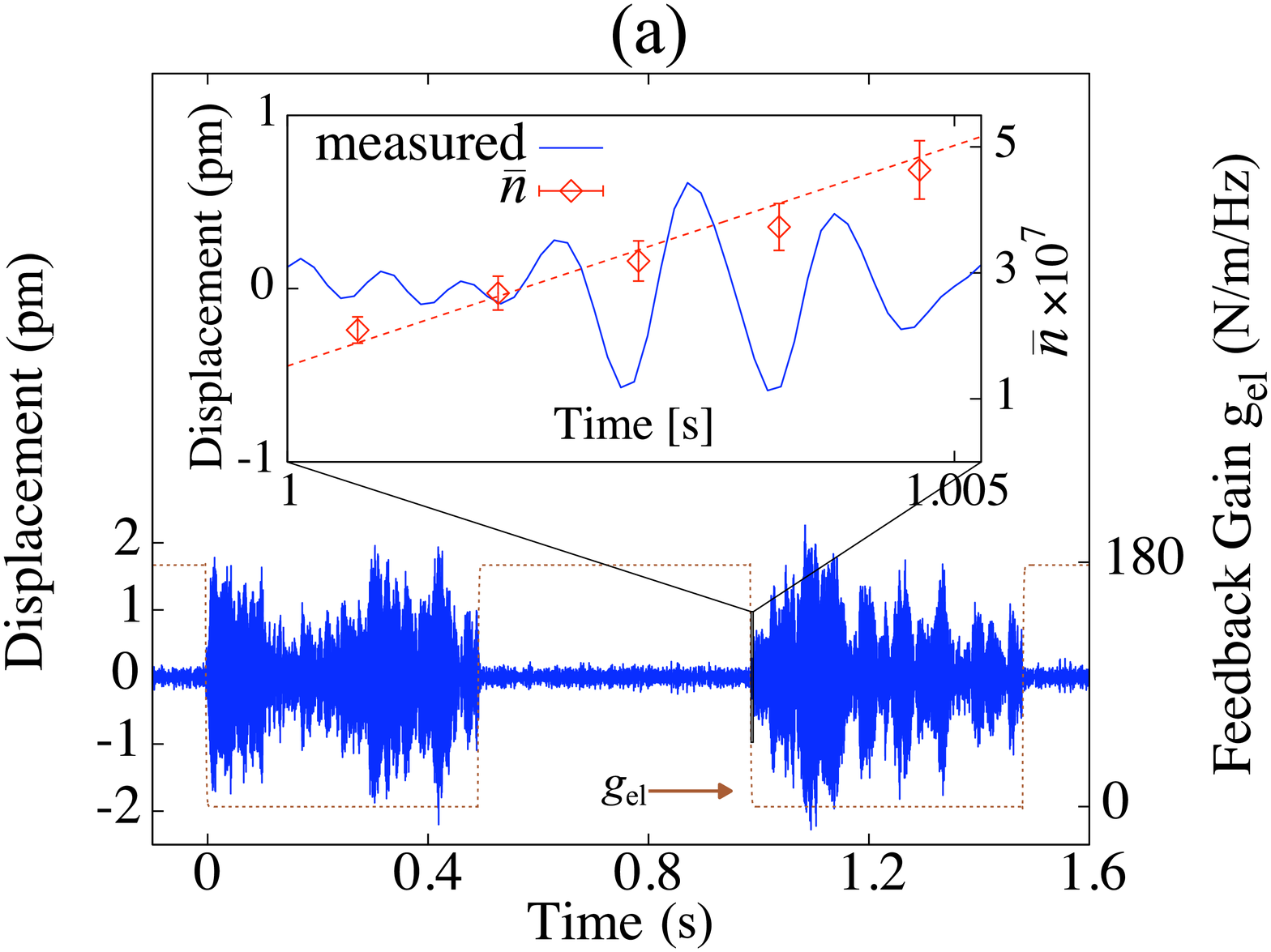}
\includegraphics[width=80mm]{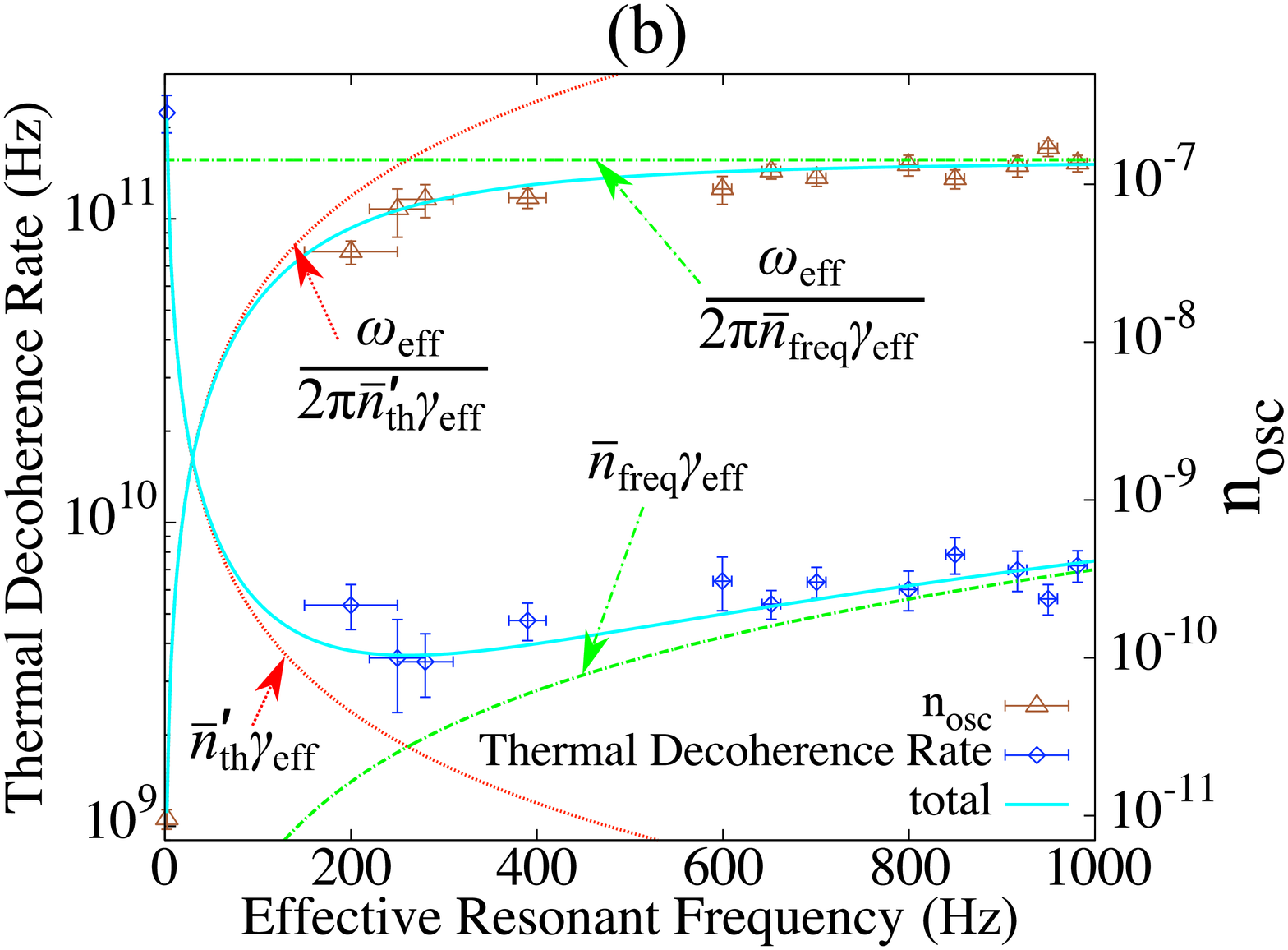}
\caption{(Color online) (a) For the case of $\omega_{\rm eff}/(2\pi)=950\ {\rm Hz}$, the measured time evolution within the first 1.6 seconds is shown. 
The inset shows an expanded region and each circle is the averaged phonon number over 100 measurements. The dotted line shows the theoretical prediction. (b) The measured decoherence rate and the corresponding value of $n_{\rm osc}$ are shown as circles and triangles, respectively. 
Here, $\bar{n}'_{\rm th}$ is $k_{\rm B}T\gamma_{\rm 1}/(\hbar\omega_{\rm eff}\gamma_{\rm eff})$, and $\bar{n}_{\rm freq}(\equiv m_{\rm 1}\omega_{\rm eff}\int_{0}^{\infty} S_x^{\rm freq}d\omega/\hbar)$ is the phonon number excited due to the optical trap. }
\label{fig4}
\end{figure}

{\it Thermal decoherence rate.}--- After cooling we rapidly make the gain of the highpass filter equal to nearly zero (as in Eq. (\ref{eq}) $m_{\rm 2}\omega_{\rm eff}^2/\kappa\sim0\ {\rm N/m/Hz}$) so that relaxation is observed. 
Since both the transfer functions $\chi_{\rm eff}$ and $\chi_{\rm eff}/(\chi_{\rm 1} g)$ consist of first-order lag elements that have the same real part of the pole $\gamma_{\rm eff}$, the average phonon number will evolve according to the expression:
\begin{eqnarray}
\frac{d\langle n\rangle}{dt}(t=0)\simeq\bar{n}_{\rm th}\frac{\omega_{\rm 1}}{\omega_{\rm eff}}\gamma_{\rm 1}+\frac{m_{\rm 1}\omega_{\rm eff}^3 S_{\rm \dot{\phi}}(\omega_{\rm eff})}{\hbar g^2}
\label{eq7}
\end{eqnarray}
Unless $\omega_{\rm eff}/(2\pi)$ exceeds $d\langle n\rangle/dt$, phonons are quickly excited so that interference due to quantum coherence is destroyed. 
Thus, $\omega_{\rm eff}/[2\pi d\langle n\rangle/dt]$ gives the oscillation number $n_{\rm osc}$ before thermal excitation. 
From Eq. (\ref{eq7}) we identify Eq. (\ref{condition}) as the relevant condition to observe the coherent oscillations of low-frequency oscillators in the presence of optical dilution. 
Note that Eq. (\ref{eq7}) is valid for not only the in-loop detector as shown in Fig. \ref{fig1}(b) but also an out-of loop detector since both signals are identical at the resonant frequency of the trapped mode.  

We use a square wave of frequency 1 Hz as an on/off switch for the cooling [Fig. \ref{fig1}(c)]. 
By injecting 100 seconds of the square wave, we repeatedly measure the time evolution of the initially cooled mode. 
We show examples of the time evolution [Fig. \ref{fig4}(a)], which represent the oscillation with random phase jumps [line in the inset of Fig. \ref{fig4}(a)].  
We also show the time evolution of the phonon number averaged over all measurements [circles in the inset of Fig. \ref{fig4}(a)], which is well fitted to the theoretical prediction [dashed line in the inset of Fig. \ref{fig4}(a)]. 
The thermal decoherence rate can be determined from the plot of the averaged phonon number versus time by linear fitting. 
We observed the rate as the cavity detuning (i.e. optical rigidity) was varied [circles in Fig. \ref{fig4}(b)].  
The results are well fitted to the theoretical prediction shown as a solid curve. 
The trap-induced decoherence presently limits the achievable rate. 
The lowest rate of $(3.5\pm0.8)\times10^9$ Hz is 60 times smaller than the bare value given by $\bar{n}_{\rm th}\gamma_1$ \cite{comment}. 
Since the optical potential dilutes energy dissipation by enhancing rigidity, the coherent oscillation number of the trapped mode is further improved by a factor of $\omega_{\rm eff}/\omega_{\rm 1}$. 
Thus, the improvement for $n_{\rm osc}$ is $10^4$-fold [triangles in Fig. \ref{fig4}(b)]. 

For the feasible case of $\omega_{\rm laser}/(2\pi)\simeq 300$ THz, $\omega_{\rm 1}/(2\pi)=1$ Hz, and $Q_{\rm 1}=5\times10^7$ \cite{cagnoli2000}, $1/n_{\rm osc}$ is equal to the following expression from Eq. (\ref{eq7}): 
\begin{eqnarray}
0.8\left(\frac{1\ {\rm kHz}}{\frac{\omega_{\rm eff}}{2\pi}}\right)^2\!\!\!+0.1\frac{m_{\rm 1}}{5{\rm \ mg}}\left(\frac{S_{\rm \dot{\phi}}^{1/2}(\omega_{\rm eff})}{4{\rm \ \frac{mHz}{\sqrt{Hz}}}}\frac{\frac{\omega_{\rm eff}}{2\pi}}{1{\rm \ kHz}}\frac{L}{5{\rm \ cm}}\right )^2
\end{eqnarray}
Here the value of $S_{\rm \dot{\phi}}$ is also feasible by stabilization \cite{numata2004}. 
$n_{\rm osc}>1$ is thus achievable by optical dilution even in the presence of the optical-trap induced decoherence. 

{\it Conclusion.}---
We directly measured the thermal decoherence rate of an optically trapped pendulum mode as the optical rigidity was varied. 
We demonstrated optical dilution effect and achieved 60-fold reduction of the rate from its bare value. 
In our present system, reduction of decoherence is limited by the optical-trap induced decoherence. 
However, our analysis substantiates the feasibility of reducing the rate under the pendulum's resonant frequency within the state of the art technology. 
One fascinating prospect is that combination with the two optomechanical systems, both of which satisfy Eq. (\ref{condition}), leads to generate the entangled state of massive mirrors \cite{muller2008}. 
Our results represent an important step towards the goal of realizing quantum massive oscillators. 


{\it Acknowledgment.}--- 
We thank Mark Sadgrove for help with the manuscript and stimulating discussions. 
This work was supported by PRESTO, JST, JSPS KAKENHI Grant No. 15617498 and No. 15617499, NINS (National Institutes of Natural Sciences) Program for Cross-Disciplinary Study, Matsuo Academic Foundation, and the Grants-in-Aid for JSPS Fellows No. 27-7404. 


\begin{thebibliography}{99}
\bibitem{comment} In the limit of the zero detuning, the optical spring effect is vanished such that $S_{x}^{\rm freq}$ becomes $S_{\dot{\phi}}/g^2$. 
This represents the position sensing noise resulting from the frequency fluctuations of the laser. 
Because the frequency fluctuations does not disturb the mirror's position in this case, it does not increase the thermal  decoherence rate. 
Mathematically, this is because that the transfer function $1/g^2$ does not include the first-order lag element. 
Note that our analysis in Eq. (\ref{eq7}) is based on the assumption that the optical rigidity is larger than the mechanical rigidity. 
\end{thebibliography}
\end{document}